\documentclass{article}
\usepackage[utf8]{inputenc}

\usepackage{newpxtext,newpxmath}
\usepackage[margin=1.25in]{geometry}

\usepackage{graphicx}
\usepackage{physics}
\usepackage[disable]{todonotes}
\usepackage{caption}
\usepackage{subcaption}
\usepackage{multirow}
\usepackage{tikz}
\usepackage{hyperref}
\usepackage{biblatex}
\usepackage{enumitem,amssymb}
\newlist{todolist}{itemize}{2}
\setlist[todolist]{label=$\square$}
\usepackage{pifont}

\usepackage{fancyhdr}

\title{Classically-Boosted Variational Quantum Eigensolver}
\author{Maxwell D. Radin \and Peter Johnson}
\date{Zapata Computing, Inc., 100 Federal St., Boston, MA 02110, USA}

\begin{document}

\maketitle

\abstract{
The ability of near-term quantum computers to represent classically-intractable quantum states has brought much interest in using such devices for estimating the ground and excited state energies of fermionic Hamiltonians.
The usefulness of such near-term techniques, generally based on the Variational Quantum Eigensolver (VQE), however, is limited by device noise and the need to perform many circuit repetitions.
This paper addresses these challenges by generalizing VQE to consider wavefunctions in a subspace spanned by classically tractable states and states that can be prepared on a quantum computer.
The manuscript shows how the ground and excited state energies can be estimated using such ``classical-boosting'' and how this approach can be combined with VQE Hamiltonian decomposition techniques.
Unlike existing VQE approaches, the sensitivity to sampling error and device noise approaches zero in the limit where the classically tractable states are able to describe an eigenstate.
A detailed analysis of the measurement requirements in the simplest case, where a single computational basis state is used to boost conventional VQE, shows that the ground-state energy estimation of several closed-shell homonuclear diatomic molecules can be accelerated by a factor of approximately 10-1000.
The analysis also shows that the measurement reduction of such single basis state boosting, relative to conventional VQE, can be estimated using only the overlap between the ground state and the computational basis state used for boosting.
}

\section{Introduction}
Due to recent advances in gate-model quantum computing, there has been much interest in using near-term quantum computers for modeling electronic structure. 
This interest has largely focused on the estimation of the ground state energy, mainly through the variational quantum eigensolver (VQE) \cite{Peruzzo2014,cao2019quantum}.

Despite great interest in VQE, its utility is limited by the error introduced by device noise as well as the large number of circuit repetitions required to achieve sufficient precision in energy estimation.
Error from device noise is exacerbated by the large numbers of gates required by useful problem instances.
For example, K{\"u}hn et al. found that millions of two-qubit gates may be required to reach chemical accuracy for reactions involving small molecules \cite{Kuehn2019}.
Even if devices had sufficient fidelity to perform such a large number of operations coherently, Gonthier et al. \cite{gonthier2020} found that VQE, as it is conventionally formulated, would still be unlikely to yield an advantage in run time over classical algorithms for typical computational chemistry problems.
Advanced estimation techniques such as Quantum Phase Estimation \cite{aspuru2005simulated} and enhanced sampling \cite{wang2021minimizing} can dramatically improve the run time, but only with the availability of high fidelity quantum devices.

A number of studies have recently proposed improvements to VQE based on subspace expansion methods \cite{McClean2017,McClean2020,Takeshita2020,Urbanek2020,Manrique2021}.
These methods consider wavefunctions in a subspace derived from a quantum state that can be prepared on a quantum computer.
For example, McClean et al. proposed a method considering the subspace spanned by excitation operators applied to the state prepared on a quantum computer and found that this approach could mitigate errors from device noise \cite{McClean2017,McClean2020}.
This was extended to the virtual quantum subspace expansion, wherein the inclusion of excitations into virtual orbitals allows for a reduction in the number of required qubits \cite{Takeshita2020,Urbanek2020}.
The translation quantum-subspace expansion instead considers the application of translation operators to the state prepared on the quantum computer and can reduce the qubit and fidelity requirements for estimating the ground-state energy of periodic Hamiltonians \cite{Manrique2021}.

In addition to these subspace expansion methods, a number of recent studies have proposed techniques to leverage classical approximations to electronic structure problems to enhance VQE.
For example, Kottman et al. explored the use of multiresolution analysis \cite{kottmann2021reducing} to reduce the qubit requirements by optimizing the corresponding orbitals.
Kirby et al. proposed an approach to obtain a correction to classical estimation of the ground-state energy by using VQE to search the Hamiltonian's contextual subspace \cite{kirby2019contextuality, kirby2020classical, kirby2021contextual}.
This technique was found to significantly reduce the qubit requirements and number of Hamiltonian terms to be evaluated on the quantum computer when estimating the ground state energy of small molecules to within chemical accuracy.

While these studies show that such techniques (as well as other recent proposals such as entanglement forging \cite{eddins2021}) can significantly reduce the qubit and fidelity requirements fro VQE, there is, to our knowledge, no evidence that any proposed near-term quantum algorithm could yield an advantage over classical electronic structure methods.
Additionally, few proposals have considered how VQE might augment, rather than replace, commonly used classical electronic structure methods.
(While contextual subspace VQE uses VQE to augment a classical energy estimate, the classical estimate is derived from a non-contextual approximation to the Hamiltonian rather than electronic structure methods commonly used in computational chemistry.)

As a step towards addressing this challenge, this study introduces classically-boosted VQE (CB-VQE), a generalization of VQE that allows one to reduce the fidelity and measurement requirements by taking advantage of classical approximate solutions to an electronic structure problem.
This approach, building on prior subspace methods, considers the subspace spanned by a set of states of which some are classically tractable and others that can be prepared on a quantum computer.
This work derives the CB-VQE method, assesses the sensitivity to device and sampling error, and presents a numerical and analytical estimation of measurement requirements for the simplest version of CB-VQE.

The analysis shows that the sensitivity of estimated eigenvalues to device and sampling error can vanish in the limit that the classically tractable states approach the exact solution.
In the simplest version of classical boosting, where a single Slater determinant is used to boost conventional VQE, the number of measurements needed to estimate the ground-state energy of homonuclear closed-shell diatomic molecules is reduced by several orders of magnitude.
An analytical expression for this speedup in terms of only the overlap between the exact ground state and the Slater determinant used for boosting is derived, allowing for the utility of single-determinant classical boosting to be readily assessed for larger systems.

\section{Theory}

This section first introduces a generalized subspace expansion, wherein a quantum computer is used to find the minimum energy within a subspace spanned by states $\left\{\ket{\Phi_\alpha}\right\}$.
Next, a sensitivity analysis explores the behavior when some of the states in this set are classically tractable.
In the regime where the classically tractable states provide a good approximation of an eigenstate, the sensitivity to errors in quantum estimation is found to vanish under reasonable assumptions.

The ground state energy (as well as excited state energies) are estimated by solving the generalized eigenvalue problem \cite{McClean2017}
\begin{align}
    \label{eqn:generalized-eigenvalue-problem}
    \overline{H}\vec{v}=\lambda\overline{S}\vec{v},
\end{align}
where
\begin{align}
    [\overline{H}]_{\alpha, \beta} = H_{\alpha, \beta} &= \langle \Phi_\alpha |H|\Phi_\beta\rangle\\
    [\overline{S}]_{\alpha, \beta}=S_{\alpha, \beta} &= \langle \Phi_\alpha |\Phi_\beta\rangle,
\end{align}
where $\alpha$ and $\beta$ run over the states chosen to span the subspace of interest.

\begin{figure}
     \centering
     \begin{subfigure}[b]{0.45\textwidth}
        \centering
        \begin{tikzpicture}[scale=1.000000,x=1pt,y=1pt]
\filldraw[color=white] (0.000000, -7.500000) rectangle (148.000000, 22.500000);
\draw[color=black] (0.000000,15.000000) -- (136.000000,15.000000);
\draw[color=black] (136.000000,14.500000) -- (148.000000,14.500000);
\draw[color=black] (136.000000,15.500000) -- (148.000000,15.500000);
\draw[color=black] (0.000000,15.000000) node[left] {$\ket{0}$};
\draw[color=black] (0.000000,0.000000) -- (148.000000,0.000000);
\draw[color=black] (0.000000,0.000000) node[left] {$\ket{0^N}$};
\begin{scope}
\draw[fill=white] (12.000000, 15.000000) +(-45.000000:8.485281pt and 8.485281pt) -- +(45.000000:8.485281pt and 8.485281pt) -- +(135.000000:8.485281pt and 8.485281pt) -- +(225.000000:8.485281pt and 8.485281pt) -- cycle;
\clip (12.000000, 15.000000) +(-45.000000:8.485281pt and 8.485281pt) -- +(45.000000:8.485281pt and 8.485281pt) -- +(135.000000:8.485281pt and 8.485281pt) -- +(225.000000:8.485281pt and 8.485281pt) -- cycle;
\draw (12.000000, 15.000000) node {$H$};
\end{scope}
\draw (8.000000, -6.000000) -- (16.000000, 6.000000);
\begin{scope}
\draw[fill=white] (36.000000, -0.000000) +(-45.000000:8.485281pt and 8.485281pt) -- +(45.000000:8.485281pt and 8.485281pt) -- +(135.000000:8.485281pt and 8.485281pt) -- +(225.000000:8.485281pt and 8.485281pt) -- cycle;
\clip (36.000000, -0.000000) +(-45.000000:8.485281pt and 8.485281pt) -- +(45.000000:8.485281pt and 8.485281pt) -- +(135.000000:8.485281pt and 8.485281pt) -- +(225.000000:8.485281pt and 8.485281pt) -- cycle;
\draw (36.000000, -0.000000) node {$W$};
\end{scope}
\draw (74.000000,15.000000) -- (74.000000,0.000000);
\begin{scope}
\draw[fill=white] (74.000000, -0.000000) +(-45.000000:28.284271pt and 12.727922pt) -- +(45.000000:28.284271pt and 12.727922pt) -- +(135.000000:28.284271pt and 12.727922pt) -- +(225.000000:28.284271pt and 12.727922pt) -- cycle;
\clip (74.000000, -0.000000) +(-45.000000:28.284271pt and 12.727922pt) -- +(45.000000:28.284271pt and 12.727922pt) -- +(135.000000:28.284271pt and 12.727922pt) -- +(225.000000:28.284271pt and 12.727922pt) -- cycle;
\draw (74.000000, -0.000000) node {$V_\alpha^\dagger h_\ell V_\beta$};
\end{scope}
\filldraw (74.000000, 15.000000) circle(1.500000pt);
\begin{scope}
\draw[fill=white] (112.000000, 15.000000) +(-45.000000:8.485281pt and 8.485281pt) -- +(45.000000:8.485281pt and 8.485281pt) -- +(135.000000:8.485281pt and 8.485281pt) -- +(225.000000:8.485281pt and 8.485281pt) -- cycle;
\clip (112.000000, 15.000000) +(-45.000000:8.485281pt and 8.485281pt) -- +(45.000000:8.485281pt and 8.485281pt) -- +(135.000000:8.485281pt and 8.485281pt) -- +(225.000000:8.485281pt and 8.485281pt) -- cycle;
\draw (112.000000, 15.000000) node {$H$};
\end{scope}
\draw[fill=white] (130.000000, 9.000000) rectangle (142.000000, 21.000000);
\draw[very thin] (136.000000, 15.600000) arc (90:150:6.000000pt);
\draw[very thin] (136.000000, 15.600000) arc (90:30:6.000000pt);
\draw[->,>=stealth] (136.000000, 9.600000) -- +(80:10.392305pt);
\end{tikzpicture}
        \caption{}
        \label{fig:hadamard-matrix-element}
     \end{subfigure}
     \hfill
     \begin{subfigure}[b]{0.45\textwidth}
        \centering
        \begin{tikzpicture}[scale=1.000000,x=1pt,y=1pt]
\filldraw[color=white] (0.000000, -7.500000) rectangle (138.000000, 22.500000);
\draw[color=black] (0.000000,15.000000) -- (126.000000,15.000000);
\draw[color=black] (126.000000,14.500000) -- (138.000000,14.500000);
\draw[color=black] (126.000000,15.500000) -- (138.000000,15.500000);
\draw[color=black] (0.000000,15.000000) node[left] {$\ket{0}$};
\draw[color=black] (0.000000,0.000000) -- (138.000000,0.000000);
\draw[color=black] (0.000000,0.000000) node[left] {$\ket{0^N}$};
\begin{scope}
\draw[fill=white] (12.000000, 15.000000) +(-45.000000:8.485281pt and 8.485281pt) -- +(45.000000:8.485281pt and 8.485281pt) -- +(135.000000:8.485281pt and 8.485281pt) -- +(225.000000:8.485281pt and 8.485281pt) -- cycle;
\clip (12.000000, 15.000000) +(-45.000000:8.485281pt and 8.485281pt) -- +(45.000000:8.485281pt and 8.485281pt) -- +(135.000000:8.485281pt and 8.485281pt) -- +(225.000000:8.485281pt and 8.485281pt) -- cycle;
\draw (12.000000, 15.000000) node {$H$};
\end{scope}
\draw (8.000000, -6.000000) -- (16.000000, 6.000000);
\begin{scope}
\draw[fill=white] (36.000000, -0.000000) +(-45.000000:8.485281pt and 8.485281pt) -- +(45.000000:8.485281pt and 8.485281pt) -- +(135.000000:8.485281pt and 8.485281pt) -- +(225.000000:8.485281pt and 8.485281pt) -- cycle;
\clip (36.000000, -0.000000) +(-45.000000:8.485281pt and 8.485281pt) -- +(45.000000:8.485281pt and 8.485281pt) -- +(135.000000:8.485281pt and 8.485281pt) -- +(225.000000:8.485281pt and 8.485281pt) -- cycle;
\draw (36.000000, -0.000000) node {$W$};
\end{scope}
\draw (69.000000,15.000000) -- (69.000000,0.000000);
\begin{scope}
\draw[fill=white] (69.000000, -0.000000) +(-45.000000:21.213203pt and 12.727922pt) -- +(45.000000:21.213203pt and 12.727922pt) -- +(135.000000:21.213203pt and 12.727922pt) -- +(225.000000:21.213203pt and 12.727922pt) -- cycle;
\clip (69.000000, -0.000000) +(-45.000000:21.213203pt and 12.727922pt) -- +(45.000000:21.213203pt and 12.727922pt) -- +(135.000000:21.213203pt and 12.727922pt) -- +(225.000000:21.213203pt and 12.727922pt) -- cycle;
\draw (69.000000, -0.000000) node {$V_\alpha^\dagger V_\beta$};
\end{scope}
\filldraw (69.000000, 15.000000) circle(1.500000pt);
\begin{scope}
\draw[fill=white] (102.000000, 15.000000) +(-45.000000:8.485281pt and 8.485281pt) -- +(45.000000:8.485281pt and 8.485281pt) -- +(135.000000:8.485281pt and 8.485281pt) -- +(225.000000:8.485281pt and 8.485281pt) -- cycle;
\clip (102.000000, 15.000000) +(-45.000000:8.485281pt and 8.485281pt) -- +(45.000000:8.485281pt and 8.485281pt) -- +(135.000000:8.485281pt and 8.485281pt) -- +(225.000000:8.485281pt and 8.485281pt) -- cycle;
\draw (102.000000, 15.000000) node {$H$};
\end{scope}
\draw[fill=white] (120.000000, 9.000000) rectangle (132.000000, 21.000000);
\draw[very thin] (126.000000, 15.600000) arc (90:150:6.000000pt);
\draw[very thin] (126.000000, 15.600000) arc (90:30:6.000000pt);
\draw[->,>=stealth] (126.000000, 9.600000) -- +(80:10.392305pt);
\end{tikzpicture}
        \caption{}
        \label{fig:hadamard-overlap}
     \end{subfigure}
        \caption{Quantum circuits implementing the Hadamard-test for estimating (a) $\textup{Re} \left(\bra{\Phi_\alpha}H\ket{\Phi_\beta}\right)$ and (b) $\textup{Re}\left(\braket{\Phi_\alpha}{\Phi_\beta}\right)$.  }
        \label{fig:hadamard-tests}
\end{figure}
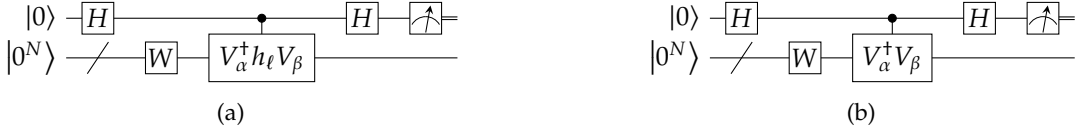

$S_{\alpha,\alpha}$ is equal to unity and $H_{\alpha,\alpha}$ can be determined using the estimation techniques used in conventional VQE \cite{cao2019quantum}.
The remaining entries are the $S_{\alpha,\beta}$ and $H_{\alpha,\beta}$ for $\alpha \ne \beta$.
To estimate these entries, let $U_\alpha$ be a unitary operator that transforms the all-zero state $\ket{0^N}$ to $\ket{\Phi_\alpha}$, i.e. $U_\alpha \ket{0^N} = \ket{\Phi_\alpha}$ where $N$ is the number of qubits.
Additionally, let $H = \sum_\ell h_\ell$ be a decomposition of the Hamiltonian into unitary operators $h_\ell$.
The off-diagonal entries can be expressed in terms of expectation values involving these operators:
\begin{equation}
    H_{\alpha,\beta} = \sum_\ell \bra{0^N} U_\alpha^\dagger h_\ell U_\beta \ket{0^N}
\end{equation}
and
\begin{equation}
    S_{\alpha,\beta} = \bra{0^N} U_\alpha^\dagger U_\beta \ket{0^N}
\end{equation}
The real and imaginary parts of these expressions can be estimated using the Hadamard test.
Fig. \ref{fig:hadamard-tests}  shows the circuits used to estimate the real parts with the state preparation unitaries for $\ket{\Phi_\alpha}$ and $\ket{\Phi_\beta}$ factorized as $U_\alpha = W_{\alpha,\beta} V_\alpha$ and $U_\beta = W_{\alpha,\beta} V_\beta$ to take advantage of common operations $W_{\alpha,\beta}$ at the beginning of the state preparations.

One strategy for implementing controlled versions of the ansatz operations $V_\alpha$ and $V_\beta$ is to promote each elementary gate of the original compilation into a controlled version of the gate \cite{Izmaylov2020}.
This introduces an additional cost to the circuit in terms of depth and number of gates.
As discussed in \ref{sec:ctrlcost}, some simplifications allow for an efficient implementation of these controls for ansatzes based on particle-number conserving blocks.
Under reasonable assumptions, these simplifications allow a control to be applied to $V_\alpha$ by increasing the number of two-qubit gates by a factor of only $1+5/N$, where $N$ is the number of qubits.

The outcome likelihoods of the circuits shown in Figure \ref{fig:hadamard-tests} are
\begin{align}
    \label{eqn:hadamard-test-hamiltonian}
    \textup{Pr}(\pm)=\frac{1}{2}\left(1\pm \textup{Re}\left(\bra{0^N} U_\alpha^\dagger h_\ell U_\beta \ket{0^N}\right)\right)
\end{align}
for the Hamiltonian entries and
\begin{align}
    \label{eqn:hadamard-test-overlap}
    \textup{Pr}(\pm)=\frac{1}{2}\left(1\pm \textup{Re}\left(\bra{0^N} U_\alpha^\dagger U_\beta \ket{0^N}\right)\right)
\end{align}
for the overlap entries.
Measurement samples give a statistical estimate of the quantities of interest 
$\textup{Re}\left(\bra{0^N} U_\alpha^\dagger h_\ell U_\beta \ket{0^N}\right)$
and
$\textup{Re}\left(\bra{0^N} U_\alpha^\dagger U_\beta \ket{0^N}\right)$.
The imaginary component of the overlap can be estimated by instead initializing the ancilla qubit in the state $\ket{-i}\ket{0^N}$, in which case the $\textup{Re}$ in Eqs. \ref{eqn:hadamard-test-hamiltonian} and \ref{eqn:hadamard-test-overlap} is replaced with $\textup{Im}$.
However, as discussed in Appendix \ref{sec:imaginary-component}, in typical cases estimating the imaginary components of $\overline{H}$ and $\overline{S}$ is not necessary.

This estimation, in both the real and imaginary cases, can be accelerated using quantum amplitude estimation or enhanced sampling methods \cite{wang2021minimizing}.
The estimation process described above yields a root mean squared error in the (unbiased) estimate of $O(1/\sqrt{M})$ where $M$ is the number of samples.
For devices with sufficiently low error, quantum amplitude estimation can improve this to $O(1/M)$.
Enhanced sampling methods, in contrast, achieve a scaling of $O(\sqrt{r/M})$ where $r$ is proportional to the error rate of elementary gates of the device.

It may be advantageous to choose some of the $\left\{\ket{\Phi_\alpha}\right\}$ basis states to be classically tractable.
Here, a set of states is said to be classically tractable if $H_{\alpha, \beta} = \langle \Phi_\alpha |H|\Phi_\beta\rangle$ and $S_{\alpha, \beta} = \langle \Phi_\alpha |\Phi_\beta\rangle$ can be calculated classically for all $\ket{\Phi_\alpha}$ and $\ket{\Phi_\beta}$ in the set.
Below, the states for which $H_{\alpha, \beta}$ and $S_{\alpha, \beta}$ are to be calculated classically are referred to as the ``classical'' states, and the remaining states in the basis as the ``quantum'' states.

The sensitivity of the generalized eigenvalues to errors in the entries of $\overline{H}$ and $\overline{S}$ can give insight into how such ``classical boosting'' might reduce the measurement and fidelity requirements.
From Eq. 12 of Ref. \cite{fox68}, \todo{is there a more textbook citation?} the sensitivity of 
a generalized eigenvalue
$\lambda$ to $H_{\alpha,\beta}$ and $S_{\alpha,\beta}$ is
\begin{equation}
    \label{eqn:sensitivity-H}
    \frac{\partial \lambda}{\partial H_{\alpha,\beta}} = v_\alpha v_\beta \left(2 - \delta_{\alpha, \beta}\right)
\end{equation}
and
\begin{equation}
    \label{eqn:sensitivity-S}
    \frac{\partial \lambda}{\partial S_{\alpha,\beta}} = - \lambda v_\alpha v_\beta \left(2 - \delta_{\alpha, \beta}\right)
\end{equation}
where $\vec{v}$ is taken to be normalized so that $\vec{v}^T \overline{S} \vec{v} = 1$ and imaginary components of $\overline{H}$ and $\overline{S}$ are neglected, as discussed above.
\todo{Why is this minus sign missing from the wikipedia article?}
(Note that the factor of two arises from taking advantage of the fact that $\textup{Re} \left(\overline{H}\right)$ and $\textup{Re}\left(\overline{S}\right)$ are symmetric.)

Suppose that the states $\left\{\ket{\Phi_\alpha}\right\}$ are mutually orthonormal.
In this case $S$ is equal to the identity and the normalization condition on $v$ reduces to $\sum_\alpha |v_\alpha|^2 = 1$.
Consider the limit where the classical states provide a good approximation of $\vec{v}$, i.e.
\begin{equation}
    \sum_{\alpha \in C} |v_\alpha|^2 \rightarrow 1
\end{equation}
where the summation runs over indices corresponding to classical states.
In this limit, the normalization condition implies that for all quantum states $\alpha$, $v_\alpha \rightarrow 0$.
Inserting this relation into Eqs. \ref{eqn:sensitivity-H} and \ref{eqn:sensitivity-S} shows that in this limit, the sensitivity to errors in $H_{\alpha,\beta}$ and $S_{\alpha,\beta}$ approaches zero when one or both of $\alpha$ and $\beta$ corresponds to a quantum state.

This asymptotic behavior shows that choosing classical states that, in linear combination, can provide a good approximation of an eigenstate may greatly reduce the measurement and fidelity requirements for estimating the exact eigenvalue to within a desired accuracy.
Importantly, as the classical approximation approaches the exact eigenstate, the sensitivity of the estimated energy to device and sampling error vanishes.

The above analysis has assumed that the classical and quantum states are mutually orthogonal.
Approximate orthogonality could be enforced by adding a penalty proportional to the sum of square overlaps $|S_{\alpha, \beta}|^2$ to the cost function used in optimizing ansatz parameters.
Alternatively, one could design quantum circuits that, by construction, yield states that are orthogonal.
For example, Nakanishi et al. have suggested the use of mutually orthogonal input states as a method for constructing a set of mutually orthogonal entangled states \cite{Nakanishi2019}.

The above approach can be implemented with a variety of choices for the quantum states and classical states.
Choices for quantum states include any of the ansatzes already proposed for VQE \cite{cao2019quantum}, whose parameters could be variationally optimized so as to minimize the ground state energy estimated from the generalized eigenvalue problem in Eq. \ref{eqn:generalized-eigenvalue-problem}.
The simplest choice for classical states are computational basis states, which can be thought of as can be thought of as boosting VQE with a configuration interaction calculation.
The case where only a single computational basis state is included is analyzed in more detail below.
Other choices for classical states include matrix-product states with low bond order as well as quantum circuits inspired by the Lipkin-Meshkov-Glick model \cite{robbins2021benchmarking}.

CB-VQE can also be used in conjunction with many Hamiltonian decomposition $\{h_\ell\}$ previously proposed for VQE.
One choice for the unitary decomposition would be the conventional decomposition of $H$ into Pauli strings \cite{Peruzzo2014}.
Others include the decomposition into unitary operators comprised of sets of mutually anti-commuting \cite{Izmaylov2020} or commuting \cite{Yen2020} Pauli strings, as well as low-rank factorizations of the Hamiltonian yielding Pauli strings conjugated by orbital rotations \cite{OF_patent,Huggins_tens_fact}.
\todo{Mention block encoding?}

\section{Measurement analysis of single-determinant boosting}

The simplest example of the CB-VQE considers just two states, one quantum and one classically tractable, with the classically tractable state being a computational basis state.
Under common encodings, a computational basis state corresponds to a single Slater determinant.
Given that the Slater determinant that provides the best approximation to the ground state is the Hartree-Fock state, we will refer to this method as HF-VQE.
This section shows how the estimation of the eigenvalue in Eq. \ref{eqn:generalized-eigenvalue-problem} can be simplified in this case and derives the relationship between the number of measurements and precision when measurements are optimally allocated to the measurement of the entries of $\overline{H}$ and $\overline{S}$.

Let classical and quantum states $\ket{\Phi_\textup{cl}}$ and $\ket{\Phi_\textup{q}}$ correspond to indices $\alpha = 1$ and $\alpha = 2$, respectively, in $H_{\alpha,\beta}$ and $S_{\alpha,\beta}$.
$H_{1,1} = \bra{\Phi_\textup{cl}}H\ket{\Phi_\textup{cl}}$ can be trivially evaluated and $H_{2,2} = \bra{\Phi_\textup{q}}H\ket{\Phi_\textup{q}}$ estimated using existing VQE approaches.
In the evaluation of $H_{1,2} = H_{2,1}^* = \bra{\Phi_\textup{cl}}H\ket{\Phi_\textup{q}}$, the Hamiltonian can be decomposed into measurable components by inserting a resolution of the identity operator into $\bra{\Phi_\textup{cl}}H\ket{\Phi_\textup{q}}$:
\begin{equation}
    \label{eqn:projected-crossterms}
    H_{1,2} = \bra{\Phi_\textup{q}}H\ket{\Phi_\textup{cl}}
    = \sum_{i} y_i \bra{i}H\ket{i_0} 
\end{equation}
where 
$y_i = \textup{Re}(\braket{\Phi_\textup{q}}{i})$ and the sum runs over all computational basis states $i$ that the Hamiltonian couples to the computational basis state $\ket{i_0} \equiv \ket{\Phi_\textup{cl}}$.

The entries of the overlap matrix $\overline{S}$ are given by 
\begin{equation}
\label{eqn:hf-overlap-matrix}
  S_{\alpha,\beta} =
    \begin{cases}
      1 & \text{if $\alpha =\beta$}\\
      y_{i_0} & \text{otherwise}
    \end{cases}.     
\end{equation}

The overlap between $\ket{\Phi_\textup{q}}$ and a given computational basis state $\ket{i}$ can be estimated using the Hadamard test as shown in Fig. \ref{fig:hadamard-overlap} by choosing $U_\textup{cl}$ to be an operator that transforms the $\ket{0^N}$ state to the desired computational basis state.

When the generalized eigenvalue problem Eq. \ref{eqn:generalized-eigenvalue-problem} is used to estimate $\lambda$ given estimators $\left\{\hat{y}_i\right\}$ and $\hat{H}_{2,2}$ for $\left\{y_i\right\}$ and $H_{2,2}$, the rules for propagation of uncertainty show that the variance in the estimator $\hat{\lambda}$ is
\begin{equation}
    \label{eqn:general-variance-formula}
    \textup{Var}\left(\hat{\lambda}\right) \approx  \sum_i \left(\frac{\partial \lambda}{\partial y_i}\right)^2 \textup{Var}\left(\hat{y}_i\right) + \left(\frac{\partial \lambda}{\partial H_{2,2}}\right)^2 \textup{Var}\left(\hat{H}_{2,2}\right)
\end{equation}
when the variances $\textup{Var}\left(\hat{y}_i\right)$ and $\textup{Var}\left(\hat{H}_{2,2}\right)$ are small.

The chain rule allows the derivative of $\lambda$ with respect to $y_i$ to be expressed as
\begin{equation}
    \frac{\partial \lambda}{\partial y_i} =
    \sum_{\alpha,\beta} \left( 
        \frac{\partial \lambda}{\partial H_{\alpha,\beta}} \frac{\partial H_{\alpha,\beta}} {\partial {y_i}} 
        +
        \frac{\partial \lambda}{\partial S_{\alpha,\beta}} \frac{\partial S_{\alpha,\beta}} {\partial{y_i}}
    \right).
\end{equation}
Combining this with Eqs. \ref{eqn:sensitivity-H}, \ref{eqn:sensitivity-S},
\ref{eqn:projected-crossterms}, and \ref{eqn:hf-overlap-matrix} and using the fact that $H_{1,1}$ and $H_{2,2}$ do not depend on $y_i$ yields  
\begin{equation}
 \frac{\partial \lambda}{\partial y_i} =
       2 v_1 v_2 \left(\bra{i}H\ket{i_0} 
        -
    \lambda \delta_{i, i_0}\right).
\end{equation}
Inserting the estimator relations corresponding to above expression and Eq. \ref{eqn:sensitivity-H} into \ref{eqn:general-variance-formula} shows that the variance in the eigenvalue estimator is given by
\begin{equation}
    \label{eqn:eigenvalue-variance-relation}
    \textup{Var}\left(\hat{\lambda}\right) \approx   4 v_1^2 v_2^2 \sum_i \left( \bra{i}H\ket{i_0} 
        -
     \lambda \delta_{i, i_0} \right)^2 \textup{Var}\left(\hat{y}_i\right) + v_2^4 \textup{Var}\left(\hat{H}_{2,2}\right)
\end{equation}
where the true weights $v_1$, $v_2$ and true eigenvalue $\lambda$ are used instead of the corresponding estimators because they are equal in the limit that the variances $\textup{Var}\left(\hat{y}_i\right)$ and $\textup{Var}\left(\hat{H}_{2,2}\right)$ are small.\footnote{If the eigenvalue is degenerate, then there are additional complications in attributing a sensitivity to eigenvector components. The present analysis therefore applies only if the two generalized eigenvalues are distinct, which will be true in the typical case.}

Note that in the case where  $\ket{\Phi_\textup{q}}$ is orthogonal to $\ket{\Phi_\textup{cl}}$ and when the state $v_1 \ket{\Phi_\textup{cl}} + v_2 \ket{\Phi_\textup{q}}$ corresponds to the exact ground state $\ket{\Phi_\textup{gs}}$, then $v_1 = \braket{\Phi_\textup{cl}}{\Phi_\textup{gs}} \equiv \alpha$ and $v_2 = \sqrt{1-\alpha^2}$.

To determine $\textup{Var} \left(\hat{y_i}\right)$, first rewrite Eq. \ref{eqn:hadamard-test-hamiltonian} as
\begin{align}
\label{eqn:hadamard-test-variables}
y_i =
    2 p_i - 1
\end{align}
where $p_i = \textup{Pr}(+|\braket{\Phi_\textup{q}}{i})$.
Let $\hat{p}_i$ be the estimator for $p_i$ corresponding to the sample mean of measurements on the ancilla qubit in the $\left(+,-\right)$ basis.
Eq. \ref{eqn:hadamard-test-variables} can be used to estimate $y_i$, and the corresponding estimator $\hat{y_i}$ has variance
\begin{align}
\label{eqn:overlap-variance}
\textup{Var}\left(\hat{y_i}\right) =
    4 \textup{Var} \left( \hat{p_i} \right)
\end{align}
Since $\hat{p}_i$ corresponds to the sample mean of a Bernoulli distribution, its variance is
\begin{align}
\label{eqn:bernoulli-variance}
\textup{Var} \left(\hat{p}_i\right) = \frac{p_i\left(1-p_i\right)}{M_i}
\end{align}
where $M_i$ is the number of samples used to estimate $p_i$. Combining Eqs. \ref{eqn:hadamard-test-variables}, \ref{eqn:overlap-variance}, and \ref{eqn:bernoulli-variance} allows the variance of the overlap estimator $\hat{y}_i$ to be expressed in terms of the overlap $y_i$:
\begin{align}
    \label{eqn:overlap-variances-solved}
    \textup{Var}\left(\hat{y}_i\right)
    = \frac{1 - y_i^2}{M_i}
\end{align}

Similarly, the measurement analysis for conventional VQE shows that if $M'$ measurements are used to estimate $H_{2,2}$, then
\begin{equation}
  \label{eqn:variance-H22}
  \textup{Var} \left( \hat{H}_{2,2} \right) = \frac{K'}{M'}
\end{equation}
where $K'$ depends on the details of the estimation technique (e.g., use of grouping or other Hamiltonian decompositions) \cite{gonthier2020, Izmaylov2020, Yen2020, OF_patent,Huggins_tens_fact}.

Previous works \cite{Rubin2018} have shown that given a set of random variables $x_j$ with estimators $\hat{x}_j$ where
\begin{equation}
\textup{Var} \left( \hat{x}_j\right) = \frac{\sigma_j^2}{N_j}
\end{equation}
then when a function $z\left(\left\{ x_i \right\} \right)$ such that $\textup{Var} \left(\hat{z}\right) = \sum_i a_i^2 \textup{Var} \left(\hat{x}_i\right)$ is to be estimated subject to the constraint $\sum_j N_j = N$, then the optimal choice of $N_j$ yields a variance of
\begin{align}
    \label{eqn:general-optimal-allocation}
    \textup{Var} \left( \hat{z} \right) = \frac{\left( \sum_j \left|a_j\right| \sigma_j  \right)^2}{N}.
\end{align}
This relation applied to Eqs. \ref{eqn:eigenvalue-variance-relation}, \ref{eqn:overlap-variances-solved}, and \ref{eqn:variance-H22}, yields
\begin{equation}
    \label{eqn:energy-variance-K}
    \textup{Var} \left( \hat{E} \right)
    =
    \frac{K}{M}
\end{equation}
where $M = \sum_i M_i + M'$ is the total number of measurements and
\begin{equation}
    \label{eqn:hf-vqe-k}
    K = \left( 2 \alpha\sqrt{1-\alpha^2}
    \sum_i |\bra{i}H\ket{i_0}
    -
    E \delta_{i,i_0}| \sqrt{1-y_i^2} + \left(1-\alpha^2\right) \sqrt{K'} \right)^2.
\end{equation}
Eqs. \ref{eqn:energy-variance-K} and \ref{eqn:hf-vqe-k} allow one to estimate the number of measurements needed to achieve a desired precision in terms of the overlaps $y_i$ and $\alpha$ and the measurement factor $K'$.

Importantly, as the overlap $\alpha$ goes to unity, $K$ approaches zero.
This means that in the limit that the classically tractable state $\ket{\Phi_\textup{cl}}$ provides a good approximation of the ground state, the number of measurements required to reach a given precision will approach zero.

Note that in the practical implementation of classical boosting, one does not know $\alpha$, $E$, and each $y_i$ a priori.
Therefore one would use approximations for these quantities for the purposes of allocating measurements to the estimation of $H_{2,2}$ and each $y_i$.
This is analogous to conventional VQE, where one uses approximate expectation values of each Pauli to allocate measurements to groups \cite{gonthier2020}.

\section{Numerical estimates of measurement counts}
To explore the speedup provided by classical boosting, this section presents numerical calculations of measurement requirements of HF-VQE as well as conventional VQE for closed-shell homonuclear diatomic molecules.
Hamiltonians were generated using Psi4 \cite{Parrish2017} with cc-pVQZ basis sets.
Estimates use canonical orbitals, although the findings of Tubman et al. \cite{Tubman2018} suggest that HF-VQE may require fewer measurements when orbital rotations are used to obtain a Slater determinant with greater overlap with the ground state.
After selecting an active space, OpenFermion \cite{mcclean2020openfermion} was used to apply the Jordan-Wigner transformation and obtain the ground state within that active space.
This ground state was used to obtain overlaps $y_i$ as well as the term variances needed for estimating measurements for conventional VQE.
These calculations were performed using Zapata's Orquestra™ platform.

For these conventional VQE measurement estimates, co-measurable terms were grouped using a greedy algorithm.
Variances in the estimators for each term were computed using the exact ground state while covariances were neglected \cite{gonthier2020}.
The measurement estimations for classically-boosted VQE employ Eqs. \ref{eqn:energy-variance-K} and \ref{eqn:hf-vqe-k}.
This corresponds to single-determinant boosting with the quantum state orthogonal to the classical state.

\begin{figure}[ht]
    \centering
    \includegraphics[width=0.8\textwidth]{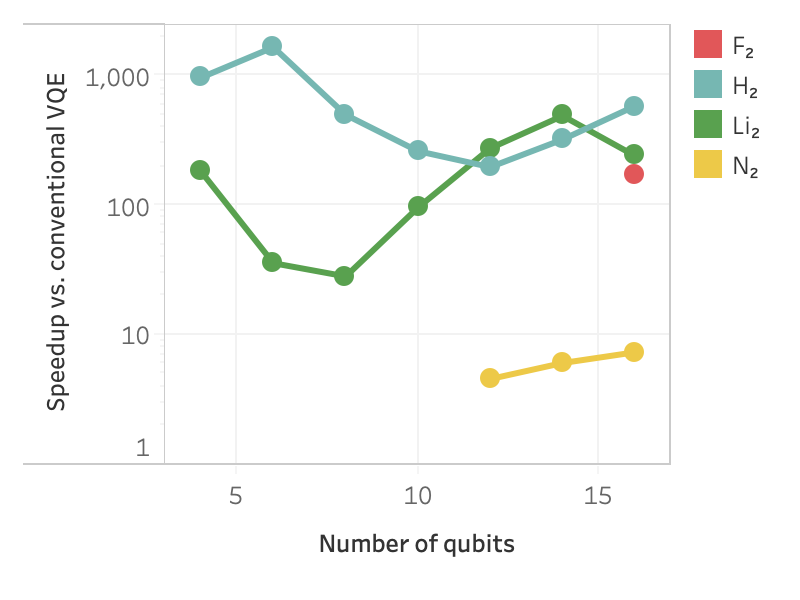}
    \caption{Speedup, in terms of number of measurements, of HF-VQE relative to conventional VQE for closed-shell homonuclear diatomic molecules.}
    \label{fig:speedup}
\end{figure}

Table \ref{table:measurement-estimates} shows the resulting estimated number of measurements to achieve a precision of 1~mHa with conventional VQE and HF-VQE.
As illustrated by Figure \ref{fig:speedup}, classical boosting reduces the number of measurements required by $\sim 1-3$ orders of magnitude for these systems.
The speedup, for a given molecule, is relatively consistent across the range of qubit counts considered, although some oscillations are present.

\begin{figure}[ht]
    \centering
    \includegraphics[width=0.8\textwidth]{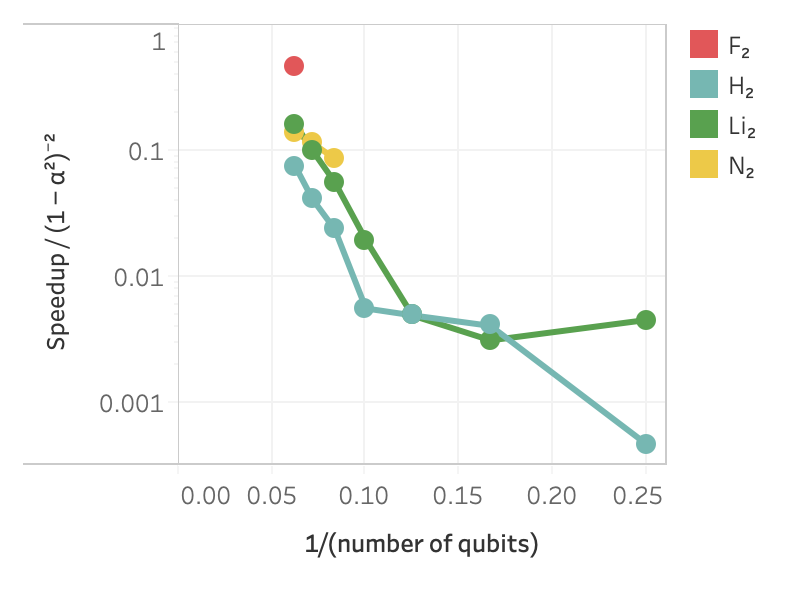}
    \caption{Speedup, in terms of number of measurements, of HF-VQE relative to conventional VQE for closed-shell homonuclear diatomic molecules, relative to the speedup expected in the asymptotic limit of large basis sets. The $S/\left(1-\alpha\right)^{-2} = 1$ position on the vertical axis corresponds to expected asymptotic limit. We find that the data trend toward this asymptotic limit, giving evidence for the validity of this prediction.}
    \label{fig:speedup-ratio}
\end{figure}

To gain greater insight into the comparison between HF-VQE and conventional VQE in large basis set (i.e. high qubit count) regime, the Appendix shows that in the limit of a large number of spin-orbitals $N_\textup{orb}$ (for fixed number of electrons), then the speedup $S$ in terms of number of measurements of HF-VQE relative to conventional VQE (when the Hamiltonian is decomposed into groups of co-measureable terms) is approximately
\begin{equation}
    \label{eqn:asymptotic-speedup}
     \lim_{N_\textup{orb} \rightarrow \infty} S \approx
     \frac{1}{\left(1-\alpha^2\right)^2}.
\end{equation}
Figure \ref{fig:speedup-ratio} shows the speedup obtained numerically relative to this asymptotic approximation.
The data shows that as the number of qubits approaches infinity, the speedup approximately aproaches that expected from Eq. \ref{eqn:asymptotic-speedup}, corresponding to the $S/\left(1-\alpha\right)^{-2} = 1$ position on the vertical axis of Figure \ref{fig:speedup-ratio}. 

\begin{table}
\centering
\begin{tabular}{cccccc}
 \hline
  Molecule & Number of qubits & Overlap & \multicolumn{2}{c}{Measurements} & Speedup\\
  & & & Conventional VQE & HF-VQE & \\ 
  \hline
  \multirow{4}{0em}{H$_2$} & 4 & 0.9997 & $3.1 \times 10^3$ & $3.2$& $970$ \\ 
  & 8 & 0.9984 &  $1.9 \times 10^5$ &$3.9 \times 10^2$ & $490$ \\
  & 12 & 0.9945& $2.7 \times 10^6$ & $1.4 \times 10^4$ & $190$ \\
  & 16 & 0.9944 & $1.7 \times 10^7$ & $2.9 \times 10^4$ & $570$ \\
  \hline
  \multirow{4}{0em}{Li$_2$} & 4 & 0.9975& $1.2 \times 10^3$ & $6.4$ & $190$ \\ 
  & 8 & 0.9934 & $7.2 \times 10^3$ & $2.5 \times 10^2$ & $29$ \\
  & 12 & 0.9928 & $2.1 \times 10^5$ & $7.6 \times 10^2$ & $280$ \\
  & 16 & 0.9870 & $1.4 \times 10^6$ & $6.0 \times 10^3$ & $230$ \\
  \hline
  \multirow{2}{0em}{N$_2$} & 12 & 0.9287 & $1.3 \times 10^7$ & $3.2 \times 10^6$ & $4.1$ \\ 
  & 16 & 0.9287 & $2.6 \times 10^7$ & $4.1 \times 10^6$ & $6.3$ \\
  \hline
  \multirow{1}{0em}{F$_2$} & 16 & 0.9737 & $1.7 \times 10^8$ & $1.0 \times 10^6$ 
 & $170$ \\
  \hline
\end{tabular}
\caption{Estimated number of measurements required to reach a precision of 1~mHa with conventional VQE and HF-VQE and the speedup (in terms of number of measurements).}
\label{table:measurement-estimates}
\end{table}

\section{Conclusions}
This paper has introduced the notion of classically-boosting VQE calculations to reduce measurement and fidelity requirements in estimating Hamiltonian eigenvalues.
The derivation of this method shows that when the classically tractable states used for boosting provide a good approximation of the desired eigenstate, then the sensitivity to sampling and device error vanishes.
Although the circuit depth required is in general greater than that required for conventional VQE by a constant factor, the lower sensitivity to device noise may nevertheless result in lower fidelity requirements.\todo{Is there a hidden assumption here about the circuit depth needed for the ansatz itself?}
Additionally, for ansatzes based on particle-number conserving blocks, classical boosting can be applied with asymptotically negligible overhead.

The simplest version of classical boosting, where a single computational basis state is used to boost a single quantum state, is found to reduce the number of required measurements by a factor of $10-1000$ for selected closed-shell homonuclear diatomic molecules.
Furthermore, the data show that the speedup can be well approximated using only the overlap between the ground state and Hartree-Fock state.

This approximation for the speedup from single-determinant boosting can be used to estimate the effectiveness of classical boosting in accelerating calculations on systems of practical interest.
FeMoco has been highlighted as an examplary system of practical interest for which quantum computers could provide value.
Using the square overlap of 0.77 found by Tubman et al. for this system \cite{Tubman2018} indicates that, in the limit of large basis size, HF-VQE would reduce the number of shots required for ground-state energy calculations on this system by a factor of approximately 19.
A greater speedup potentially could be achieved by using a classical state (or linear combination of classical states) that incorporates electron correlation.

Determining whether classically boosted VQE could outperform classical methods, while beyond the scope of this manuscript, is a crucial question for guiding the direction of future research on this topic.
However, one distinguishing feature of CB-VQE compared to other near-term quantum algorithms for electronic structure is that it uses VQE to augment, rather than replace, classical electronic structure methods.
This suggests that CB-VQE may be able to perform at least as well as the classical methods that could be used for boosting, such as Hartree-Fock, configuration interaction, or matrix-product state methods.
In order to have a more fair comparison to our method,
it would be useful to have measurement cost analyses for many of the recently proposed extensions to VQE \cite{McClean2017,McClean2020,Takeshita2020,Urbanek2020,Manrique2021,kottmann2021reducing,kirby2021contextual,eddins2021}.
The measurement costing formalism introduced in this paper may be used to assess the measurement costs of those techniques that are similarly based on generalized eigenvalue problems \cite{McClean2017,McClean2020,Takeshita2020,Urbanek2020,Manrique2021}.






\section*{Acknowledgement \label{Acknowledgement}}
The authors acknowledge insightful scientific discussions and suggestions from J\'er\^ome F. Gonthier, Alex Kunitsa, Peter Love, and Al\'{a}n Aspuru-Guzik.

\printbibliography

\appendix 

\section{Imaginary components of the generalized eigenvalue problem}
\label{sec:imaginary-component}
Here we show that ignoring the imaginary components of $\overline{H}$ and $\overline{S}$ still results in a valid ground state energy upper bound.
Consider the generalized eigenvalue problem
\begin{align}
    \label{eqn:generalized-eigenvalue-problem-real}
    \textup{Re} \left(\overline{H}\right)\vec{u}=\lambda_\textup{R} \textup{Re} \left(\overline{S}\right) \vec{u}.
\end{align}
Because $\textup{Re} \left(\overline{H}\right)$ and $\textup{Re} \left(\overline{S}\right)$ are real and symmetric, the eigenvalues $\lambda_\textup{R}$ and eigenvectors $u$ are real.
Let $\lambda_\textup{R}^{\left(0\right)}$ be the lowest eigenvalue and $\vec{u}^{\left(0\right)}$ be the corresponding eigenvector.

The lowest eigenvalue of the complex generalized eigenvalue problem (Eq. \ref{eqn:generalized-eigenvalue-problem}) is the minimum of the Rayleigh quotient
\begin{equation}
    \label{eqn:rayleigh-inequality}
    \lambda^{\left(0\right)} \le R\left(\overline{H}, \overline{S}, \vec{v}\right) = 
    \frac{\vec{v}^\dagger\overline{H}\vec{v}}{\vec{v}^\dagger\overline{S}\vec{v}}.
\end{equation}
Consider now
\begin{equation}
    \label{eqn:rayleigh-u0-complex}
    R\left(\overline{H}, \overline{S}, \vec{u}^{\left(0\right)}\right) = 
    \frac{ \left(\vec{u}^{\left(0\right)}\right)^\dagger\overline{H} \vec{u}^{\left(0\right)}}{ \left(\vec{u}^{\left(0\right)}\right)^\dagger\overline{S} \vec{u}^{\left(0\right)}}.
\end{equation}
This expression can be simplified because given a Hermitian matrix $\overline{A}$ and a real vector $\vec{w}$, the imaginary component of $\overline{A}$ does not contribute to $\vec{w}^\dagger A\vec{w}$.
This can be seen by noting that $\vec{w}^\dagger A\vec{w}$ is real because $\overline{A}$ is Hermitian; and the imaginary component of each entry of $\overline{A}$ cannot contribute to the real component of $\vec{w}^\dagger A\vec{w}$ because $\vec{w}$ is real.

Therefore Eq. \ref{eqn:rayleigh-u0-complex} can be rewritten as
\begin{equation}
    \label{eqn:rayleigh}
    R\left(\overline{H}, \overline{S}, \vec{u}^{\left(0\right)}\right) = 
    \frac{ \left(\vec{u}^{\left(0\right)}\right)^\textup{T}\textup{Re}\left(\overline{H}\right) \vec{u}^{\left(0\right)}}{ \left(\vec{u}^{\left(0\right)}\right)^\textup{T}\textup{Re}\left(\overline{S}\right) \vec{u}^{\left(0\right)}}.
\end{equation}
The right hand side of this equation represents the Rayleigh quotient corresponding to the real generalized eigenvalue problem of Eq. \ref{eqn:generalized-eigenvalue-problem-real}, and so
\begin{equation}
R\left(\overline{H}, \overline{S}, \vec{u}^{\left(0\right)}\right) = \lambda_\textup{R}^{\left(0\right)}.
\end{equation}
Combining this relation with Eq. \ref{eqn:rayleigh-inequality} shows that the lowest eigenvalue of the complex generalized eigenvalue problem (Eq. \ref{eqn:generalized-eigenvalue-problem} is a lower bound on the lowest eigenvalue of the real generalized eigenvalue problem (Eq \ref{eqn:generalized-eigenvalue-problem-real}):
\begin{equation}
    \label{eqn:eigenvalue-bound}
    \lambda^{\left(0\right)} \le \lambda_\textup{R}^{\left(0\right)}.
\end{equation}
Eq. \ref{eqn:eigenvalue-bound} shows that neglecting the imaginary components of $\overline{H}$ and $\overline{S}$ will never cause the ground state energy to be underestimated.

While the neglect of the imaginary components of $\overline{H}$ and $\overline{S}$ in general will lead to the ground state energy being overestimated, for Hamiltonians with time reversal symmetry (which is the case for typical chemical systems of interest), then, given suitable choices for $\left\{\ket{\Phi_\textup{cl}^{\left(j\right)}}\right\}$ and $\left\{\ket{\Phi_\textup{q}^{\left(k\right)}}\right\}$, the exact ground state energy can nevertheless be obtained from the real generalized eigenvalue problem.
Time reversal symmetry implies that for each eigenvalue, one can construct an eigenstate wherein the coefficient of each computational basis state is real.
This ground state can be obtained as a linear combination of classical and quantum wavefunctions which also have a real coefficient for each computational basis state.
In this case, all entries in $\overline{H}$ and $\overline{S}$ will be real and so $\lambda^{\left(0\right)} = \lambda_\textup{R}^{\left(0\right)}$.

In practice, one can choose the $\left\{\ket{\Phi_\textup{cl}^{\left(j\right)}}\right\}$ to be real by construction and rely on variational optimization to find the values of $\left\{\ket{\Phi_\textup{q}^{\left(k\right)}}\right\}$ that yield the true ground state energy.
It may also be possible to design ansatzes such that the quantum states are real by construction.

\section{Efficient implementation of controlled operations}
\label{sec:ctrlcost}
This section shows that ansatzes that are comprised of an initial layer of $X$ gates to prepare a single computational basis state $\ket{i}$ followed by particle-number conserving blocks allow for efficient implementation of the controlled operations shown in Figs. \ref{fig:hadamard-matrix-element} and \ref{fig:hadamard-overlap}.
Ansatzes in this class include the hardware-efficient ansatzes based on $U_{1,\textup{ex}}$ and $U_{2,\textup{ex}}$ gates \cite{Barkoutsos2018} as well as the swap-network implementation of the $k$-UpCCGSD ansatz \cite{ogorman2019}.


Consider basis states $\ket{\Phi_\alpha}$ that can be prepared by applying the operator $V_\alpha$ to the Jordan-Wigner encoding of the Hartree-Fock state, $\ket{i_0}=\ket{0^{N-k}1^{k}}=W\ket{0}$. 
The estimation of $H_{\alpha,\beta}$ and $S_{\alpha,\beta}$ using the Hadamard test as shown in \ref{fig:hadamard-matrix-element} and \ref{fig:hadamard-overlap} requires controlled versions of the unitary $V_\alpha$ and $V_\beta$.
A straight-forward approach would promote each of the blocks $V^{(j)}_\alpha$ and the Hamiltonian components $h_\ell$ to controlled versions, leading to a constant-factor increase in the circuit depth and number of gates.

However, a the circuit depth and number gates needed to implement the overlap estimation can be significantly reduced if $V_\alpha$ can be decomposed into particle-number conserving blocks, i.e $V_\alpha=V^{(T)}_\alpha\ldots V^{(1)}_\alpha$ where $V^{(T)}_\alpha,\ldots V^{(1)}_\alpha$ are operations that conserve particle number and act on adjacent qubits.
Let $\ket{\psi}$ be an eigenstate of $V$: $V\ket{\psi}=e^{i\theta}\ket{\psi}$.
For arbitrary unitaries $A$ and $B$,
\begin{align}
    c-(AVB)|+\psi>&=\frac{1}{\sqrt{2}}\ket{0}\ket{\psi}+\frac{1}{\sqrt{2}}\ket{1}AVB\ket{\psi}\\
    &=\frac{1}{\sqrt{2}}e^{-i\theta}\ket{0}V\ket{\psi}+\frac{1}{\sqrt{2}}\ket{1}AVB\ket{\psi}\\
    &=(c-A)(R_Z(-\theta)\otimes V)(c-B)|+\psi>,
\end{align}
where $R_Z(\theta)=[[e^{i\theta},0],[0,1]]$.
That is, for any operations in a sequence of controlled-gates that preserve the initial state, we can equivalently implement them as a non-controlled operation in parallel with a $Z$-rotation on the ancilla qubit.
Of the blocks in $V_q$, the only which do not preserve the Hartree-Fock state are those which couple the filled and unfilled spin orbitals.
Blocks that act entirely on qubits initialized in $\ket{0}$ or entirely on qubits initialized in $\ket{1}$ therefore need not be promoted to controlled gates.
If the blocks are arranged in a brick-layer format with $D$ layers, one ``coupling block'' occurs on every other layer of blocks, leading to just $D/2$ blocks that require a controlled-implementation.

\begin{figure}[ht]
     \centering
    \begin{tikzpicture}[scale=1.000000,x=1pt,y=1pt]
\filldraw[color=white] (0.000000, -7.500000) rectangle (216.000000, 67.500000);
\draw[color=black] (0.000000,60.000000) -- (204.000000,60.000000);
\draw[color=black] (204.000000,59.500000) -- (216.000000,59.500000);
\draw[color=black] (204.000000,60.500000) -- (216.000000,60.500000);
\draw[color=black] (0.000000,60.000000) node[left] {$\ket{0}$};
\draw[color=black] (0.000000,45.000000) -- (216.000000,45.000000);
\draw[color=black] (0.000000,45.000000) node[left] {$\ket{0}$};
\draw[color=black] (0.000000,30.000000) -- (216.000000,30.000000);
\draw[color=black] (0.000000,30.000000) node[left] {$\ket{0}$};
\draw[color=black] (0.000000,15.000000) -- (216.000000,15.000000);
\draw[color=black] (0.000000,15.000000) node[left] {$\ket{1}$};
\draw[color=black] (0.000000,0.000000) -- (216.000000,0.000000);
\draw[color=black] (0.000000,0.000000) node[left] {$\ket{1}$};
\begin{scope}
\draw[fill=white] (12.000000, 60.000000) +(-45.000000:8.485281pt and 8.485281pt) -- +(45.000000:8.485281pt and 8.485281pt) -- +(135.000000:8.485281pt and 8.485281pt) -- +(225.000000:8.485281pt and 8.485281pt) -- cycle;
\clip (12.000000, 60.000000) +(-45.000000:8.485281pt and 8.485281pt) -- +(45.000000:8.485281pt and 8.485281pt) -- +(135.000000:8.485281pt and 8.485281pt) -- +(225.000000:8.485281pt and 8.485281pt) -- cycle;
\draw (12.000000, 60.000000) node {$H$};
\end{scope}
\draw (44.000000,60.000000) -- (44.000000,15.000000);
\begin{scope}
\draw[fill=white] (44.000000, 22.500000) +(-45.000000:19.798990pt and 19.091883pt) -- +(45.000000:19.798990pt and 19.091883pt) -- +(135.000000:19.798990pt and 19.091883pt) -- +(225.000000:19.798990pt and 19.091883pt) -- cycle;
\clip (44.000000, 22.500000) +(-45.000000:19.798990pt and 19.091883pt) -- +(45.000000:19.798990pt and 19.091883pt) -- +(135.000000:19.798990pt and 19.091883pt) -- +(225.000000:19.798990pt and 19.091883pt) -- cycle;
\draw (44.000000, 22.500000) node {$U_\textup{ent}$};
\end{scope}
\filldraw (44.000000, 60.000000) circle(1.500000pt);
\draw (84.000000,45.000000) -- (84.000000,30.000000);
\begin{scope}
\draw[fill=white] (84.000000, 37.500000) +(-45.000000:19.798990pt and 19.091883pt) -- +(45.000000:19.798990pt and 19.091883pt) -- +(135.000000:19.798990pt and 19.091883pt) -- +(225.000000:19.798990pt and 19.091883pt) -- cycle;
\clip (84.000000, 37.500000) +(-45.000000:19.798990pt and 19.091883pt) -- +(45.000000:19.798990pt and 19.091883pt) -- +(135.000000:19.798990pt and 19.091883pt) -- +(225.000000:19.798990pt and 19.091883pt) -- cycle;
\draw (84.000000, 37.500000) node {$U_\textup{ent}$};
\end{scope}
\draw (84.000000,15.000000) -- (84.000000,0.000000);
\begin{scope}
\draw[fill=white] (84.000000, 7.500000) +(-45.000000:19.798990pt and 19.091883pt) -- +(45.000000:19.798990pt and 19.091883pt) -- +(135.000000:19.798990pt and 19.091883pt) -- +(225.000000:19.798990pt and 19.091883pt) -- cycle;
\clip (84.000000, 7.500000) +(-45.000000:19.798990pt and 19.091883pt) -- +(45.000000:19.798990pt and 19.091883pt) -- +(135.000000:19.798990pt and 19.091883pt) -- +(225.000000:19.798990pt and 19.091883pt) -- cycle;
\draw (84.000000, 7.500000) node {$U_\textup{ent}$};
\end{scope}
\draw (116.000000,60.000000) -- (116.000000,0.000000);
\begin{scope}
\draw[fill=white] (116.000000, 22.500000) +(-45.000000:8.485281pt and 40.305087pt) -- +(45.000000:8.485281pt and 40.305087pt) -- +(135.000000:8.485281pt and 40.305087pt) -- +(225.000000:8.485281pt and 40.305087pt) -- cycle;
\clip (116.000000, 22.500000) +(-45.000000:8.485281pt and 40.305087pt) -- +(45.000000:8.485281pt and 40.305087pt) -- +(135.000000:8.485281pt and 40.305087pt) -- +(225.000000:8.485281pt and 40.305087pt) -- cycle;
\draw (116.000000, 22.500000) node {$h_\ell$};
\end{scope}
\filldraw (116.000000, 60.000000) circle(1.500000pt);
\begin{scope}
\draw[fill=white] (148.000000, 60.000000) +(-45.000000:19.798990pt and 8.485281pt) -- +(45.000000:19.798990pt and 8.485281pt) -- +(135.000000:19.798990pt and 8.485281pt) -- +(225.000000:19.798990pt and 8.485281pt) -- cycle;
\clip (148.000000, 60.000000) +(-45.000000:19.798990pt and 8.485281pt) -- +(45.000000:19.798990pt and 8.485281pt) -- +(135.000000:19.798990pt and 8.485281pt) -- +(225.000000:19.798990pt and 8.485281pt) -- cycle;
\draw (148.000000, 60.000000) node {$R_z\left(\theta\right)$};
\end{scope}
\begin{scope}
\draw[fill=white] (180.000000, 60.000000) +(-45.000000:8.485281pt and 8.485281pt) -- +(45.000000:8.485281pt and 8.485281pt) -- +(135.000000:8.485281pt and 8.485281pt) -- +(225.000000:8.485281pt and 8.485281pt) -- cycle;
\clip (180.000000, 60.000000) +(-45.000000:8.485281pt and 8.485281pt) -- +(45.000000:8.485281pt and 8.485281pt) -- +(135.000000:8.485281pt and 8.485281pt) -- +(225.000000:8.485281pt and 8.485281pt) -- cycle;
\draw (180.000000, 60.000000) node {$H$};
\end{scope}
\draw[fill=white] (198.000000, 54.000000) rectangle (210.000000, 66.000000);
\draw[very thin] (204.000000, 60.600000) arc (90:150:6.000000pt);
\draw[very thin] (204.000000, 60.600000) arc (90:30:6.000000pt);
\draw[->,>=stealth] (204.000000, 54.600000) -- +(80:10.392305pt);
\end{tikzpicture}
    \caption{Efficient implementation of the circuit for estimating $\bra{\Phi_\textup{q}}H\ket{\Phi_\textup{cl}}$ for an ansatz comprised of particle-number conserving blocks when the classical state is a single computational basis state.}
    \label{fig:hadamard-matrix-element-example}
\end{figure}
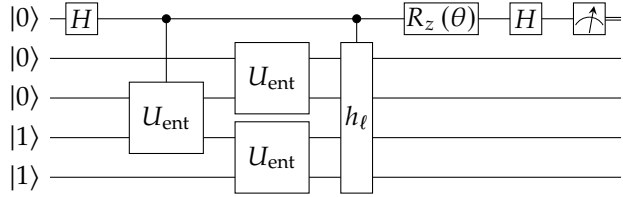

As an example, Fig. \ref{fig:hadamard-matrix-element-example} shows the circuit used to estimate $\bra{\Phi_\textup{q}}H\ket{\Phi_\textup{cl}}$ in such a case with four spin-orbitals (in an interleaved spin-orbital ordering) and two electrons where the classical state $ket{\Phi_\textup{cl}}$ is taken to be the Hartree-Fock state.
Importantly, two of the three entangling gates $U_\textup{ent}$ do not need to have controls applied.
The $R_z\left(\theta\right)$ gate applies a phase to compensate for the phase that these two gates apply in the subspace where the ancilla is in the $\ket{0}$ state.

The number of two-qubit gates needed to implement the overlap test in HF-VQE relative to the number needed for estimating an expectation value in conventional VQE can be estimated under reasonable assumptions.
Suppose that promotion of each block into a controlled version incurs a factor $F$ increase in the number of elementary gates over the original block.
Assuming that the Hamiltonian is decomposed into Pauli strings, a controlled version of each Pauli string operation $h_\ell$ can be implemented by promoting each single-qubit Pauli gate to a controlled-version of that gate, requiring at most $N$ two-qubit gates (assuming native controlled-gates and all-to-all connectivity).
The initial brick-layer arrangement of operations of $D$ layers of blocks includes approximately $D(N/2)$ blocks.
Therefore the factor increase in gates due to the controlled implementation is
\begin{align}
    \frac{DN/2+FD/2+N}{DN/2}=1+\frac{F}{N}+\frac{2}{D}.
\end{align}
Assume the ansatz is chosen have depth $D=N$. Further, assume that the blocks $V^{(j)}$ are compiled into gates with each two-qubit gate being a controlled-NOT; from \cite{barenco1995elementary}, this can be promoted to a Toffoli using three controlled-NOT gates, leading to $F\leq3$. 
Then, the total factor increase in two-qubit gates is $1+5/N$.



\section{Asymptotic speedup}
This section estimates the speedup in terms of measurements of HF-VQE relative to conventional VQE, that is
\begin{equation}
    S = \frac{M^{\left(\textup{VQE}\right)}}{M}
\end{equation}
when
\begin{equation}
    \textup{Var}\left(\hat{E}\right) = \textup{Var}\left(\hat{E}_\textup{VQE}\right),
\end{equation}
where $M^{\left(\textup{VQE}\right)}$ is the number of measurements applied to a conventional VQE ground-state energy estimator $\hat{E}_\textup{VQE}$ whose variance is given by
\begin{equation}
  \label{eqn:variance-VQE}
  \textup{Var} \left( \hat{E}_\textup{VQE} \right) = \frac{K^{\left(\textup{VQE}\right)}}{M^{\left(\textup{VQE}\right)}}.
\end{equation}
Combining these relations with Eqs. \ref{eqn:energy-variance-K} and \ref{eqn:variance-VQE} shows that $S = K^{\left(\textup{VQE}\right)}/K$, and applying Eq. \ref{eqn:hf-vqe-k} yields:
\begin{equation}
    \label{eqn:speedup-expanded}
    \frac{1}{\sqrt{S}} =  \frac{ 2 \alpha\sqrt{1-\alpha^2}
    \sum_i \bra{i}H\ket{i_0}
    -
    E \delta_{i,i_0}| \sqrt{1-y_i^2} + \left(1-\alpha^2\right) \sqrt{K'} }{\sqrt{K^{\left(\textup{VQE}\right)}}}.
\end{equation}

This expression can be simplified when the conventional VQE energy $E_\textup{VQE} = \bra{\Phi} H \ket{\Phi}$ as well as $H_{2,2} = \bra{\Phi_\textup{q}} H \ket{\Phi_\textup{q}}$ are estimated by the independent measurement of Pauli terms in the Hamiltonian.
In this case,
\begin{equation}
    K^{\left(\textup{VQE}\right)} = \left(  \sum_{i} |h_i| \sqrt{\textup{Var}_{\ket{\Phi}} \left( \hat{P}_i \right)  } \right)^2
\end{equation}
where $\hat{P}_i$ is the estimator for the $i^\textup{th}$ Pauli term in the Hamiltonian and $h_i$ is its coefficient \cite{gonthier2020}.
Similarly,
\begin{equation}
    K' = \left(  \sum_{i} |h_i| \sqrt{\textup{Var}_{\ket{\Phi_\textup{q}}} \left( \hat{P}_i \right)  } \right)^2
\end{equation}
These quantities can be approximated by setting the variances to their upper bound of unity, in which case
\begin{equation}
K^{\left(\textup{VQE}\right)}  \approx K' \approx \left(\sum_i |h_i|\right)^2.
\end{equation}
Prior studies have found that this approximation yields good estimates for VQE measurement requirements \cite{gonthier2020}.
One can analogously approximate $\sqrt{1 - y_i^2}$ with the upper bound of unity in Eq. \ref{eqn:speedup-expanded}.

Applying these approximations to Eq. \ref{eqn:speedup-expanded} yields a simplified expression for the speedup:
\begin{equation}
\label{eqn:speedup-partly-simplified}
    \frac{1}{\sqrt{S}} =  \frac{ 2 \alpha\sqrt{1-\alpha^2}
    \sum_i |\bra{i}H\ket{i_0}
    -
    E \delta_{i,i_0}|   }{\sqrt{K^{\left(\textup{VQE}\right)}}} + \left(1-\alpha^2\right).
\end{equation}

Because each Pauli string $P_i$ that does not annihilate $\ket{i_0}$ will transform it to a single computational basis state,
\begin{equation}
    \sum_{i} |\bra{i}H\ket{i_0}| \le \sum_{i \in T}|h_j|
\end{equation}
where $T$ is the set of Pauli terms that do not annihilate $\ket{i_0}$.

For typical electronic structure problems, $T$ includes $O\left(N_\textup{orb}^2 N_\textup{el}^2\right)$ terms, where $N_\textup{orb}$ is the number of spin-orbitals included in the calculation and $N_\textup{el}$ the number of electrons.
In contrast, the full Hamiltonian contains $O\left(N_\textup{orb}^4\right)$ terms.
Therefore in the limit of large $N_\textup{orb}$ (with $N_\textup{el}$ fixed), the ratio $ \sum_{j \in T}|h_j|/ \sum_j|h_j|$ will vanish provided that the ratio of the mean absolute coefficient in $T$ to the mean absolute coefficient of the full Hamiltonian scales subquadratically in $N_\textup{orb}$.
In such cases,
\begin{equation}
     \lim_{N_\textup{orb} \rightarrow \infty} 
    \frac {\sum_i| \bra{i}H\ket{i_0}
    -
    E \delta_{i,i_0}| }{\sum_i |h_i|}
   =
    \frac {|\bra{i_0}H\ket{i_0}-E| - |\bra{i_0}H\ket{i_0}| }{\sum_i |h_i|}
\end{equation}

Note also that for typical electronic structure problems, the numerator on the right-hand side of the above equation will approach a constant in the $N_\textup{orb} \rightarrow \infty$ limit, while the denominator will diverge.
Therefore 
\begin{equation}
     \lim_{N_\textup{orb} \rightarrow \infty} 
    \frac {\sum_i| \bra{i}H\ket{i_0}
    -
    E \delta_{i,i_0}| }{\sum_i |h_i|}
   = 0
\end{equation}
Taking the $N_\textup{orb} \rightarrow \infty$ limit of Eq. \ref{eqn:speedup-partly-simplified}, inserting the above relation, and rearranging yields

\begin{equation}
    \label{eqn:speedup-final}
     \lim_{N_\textup{orb} \rightarrow \infty} S \approx
     \frac{1}{\left(1-\alpha^2\right)^2}.
\end{equation}

Note that this derivation has assumed that the Hamiltonian is decomposed into co-measurable groups \cite{Peruzzo2014} for the purpose of estimating $H_{2,2}$ and the conventional VQE energy.
Modifications may be required when other Hamiltonian decomposition techniques are used \cite{Izmaylov2020,Yen2020,OF_patent,Huggins_tens_fact}
.
In particular, Gonthier et al. \cite{gonthier2020}, have shown that for other decomposition techniques the upper bound on variances does not provide a good approximation of the measurement requirements.
\end{document}